\documentclass[sigconf,nonacm,9pt]{acmart}

\newcommand\vldbdoi{XX.XX/XXX.XX}

\newcommand\vldbavailabilityurl{http://vldb.org/pvldb/format_vol14.html}

\usepackage[english]{babel}
\usepackage{caption}
\usepackage{subcaption}
\usepackage{algpseudocode}
\usepackage{lscape}
\usepackage[linesnumbered,ruled,vlined]{algorithm2e}
\usepackage{amsmath}
\usepackage{graphicx}
\usepackage{cleveref}
\usepackage{enumitem}
\usepackage{adjustbox}

\usepackage{multirow}

\usepackage{listings}

\usepackage{colortbl}

\usepackage{tikz}
\usetikzlibrary{shapes.arrows}

\lstdefinelanguage{Cypher}%
  {morekeywords={MATCH, OPTIONAL, WITH, SUM, RETURN},
   sensitive=true,
   morecomment=[l]{//},
   morecomment=[s]{/*}{*/},
   morestring=[b]",
  }

\newcommand{\sql}[1]{\lstinline[language=SQL]{#1}}

\newtheorem{theo}{Theorem}[section]

\newtheorem{example}[theo]{Example}

\theoremstyle{remark}

\usepackage{dsfont}

\def\HS{\hspace{\fontdimen2\font}}
\definecolor{darkgreen}{rgb}{0.15,0.55,0.15}
\definecolor{darkblue}{rgb}{0.1,0.1,0.5}
\definecolor{blue}{rgb}{0.01,0.40,.8}
\definecolor{darkgreen}{rgb}{0.15,0.55,0.15}
\definecolor{mred}{rgb}{.80,.12,.30}
\definecolor{grey}{rgb}{0.5,0.5,0.5}
\definecolor{Purple}{rgb}{.75,0,.85}
\definecolor{light-gray}{gray}{0.95}
\definecolor{mid-gray}{gray}{0.85}
\definecolor{darkred}{rgb}{0.7,0.25,0.25}
\definecolor{rose}{rgb}{1.0, 0.01, 0.24}
\newcommand{\red}[1]{\textcolor{red}{#1}}

\newcommand{\blue}[1]{\textcolor{blue}{#1}}

\usepackage{latexsym}

\DeclareRobustCommand{\ojoin}{\rule[0.10ex]{.3em}{.4pt}\llap{\rule[1.40ex]{.3em}{.4pt}}}
\newcommand{\leftouterjoin}{\mathrel{\ojoin\mkern-6.5mu\Join}}

\newcommand{\eat}[1]{}

\newcommand{\stitle}[1]{\vspace{2pt}\noindent\textbf{#1}}

\newcommand{\cnt}[0]{\emph{COUNT}\xspace}

\author{Zezhou Huang}
\email{zh2408@columbia.edu}
\affiliation{
  \institution{Columbia University}
}

\author{Pavan Kalyan Damalapati}
\email{pd2720@columbia.edu}
\affiliation{
  \institution{Columbia University}
}

\author{Eugene Wu}
\email{ewu@cs.columbia.edu}
\affiliation{
  \institution{DSI, Columbia University}
}

\begin{document}

\title{Aggregation Consistency Errors in Semantic Layers and How to Avoid Them}

\begin{abstract}
Analysts often struggle with analyzing data from multiple tables in a database due to their lack of knowledge on how to join and aggregate the data.
To address this, data engineers pre-specify "semantic layers" which include the join conditions and "metrics" of interest with aggregation functions and expressions.
However, joins can cause "aggregation consistency issues".
For example, analysts may observe inflated total revenue caused by double counting from join fanouts. 
Existing BI tools rely on heuristics for deduplication, resulting in imprecise and challenging-to-understand outcomes.
To overcome these challenges, we propose "weighing" as a core primitive to counteract join fanouts. 
"Weighing" has been used in various areas, such as market attribution and order management, ensuring metrics consistency (e.g., total revenue remains the same) even for many-to-many joins.
The idea is to assign equal weight to each join key group (rather than each tuple) and then distribute the weights among tuples. Implementing weighing techniques necessitates user input; therefore, we recommend a human-in-the-loop framework that enables users to iteratively explore different strategies and visualize the results.

\end{abstract}

\maketitle



\section{Introduction}

Cloud technology has enabled enterprises to store unlimited amounts of tables in data warehouses. 
However, analysts with domain knowledge of the business but with limited expertise in data management face challenges when analyzing metrics of interest across multiple tables. In particular, they may find it difficult to determine which tables to join, what the join conditions are, and how the attributes of these tables relate to the metrics of interest~\cite{joinconfusingtweet3,joinconfusingtweet2}.

To bridge this knowledge gap, one popular and easy approach is to denormalize (join) tables into a wide table~\cite{denorm,denorm2}, which is usually carried out by data engineers offline. 
This simplifies the data analysis process for analysts, who only need to work with one table as the single source of truth.
However, join causes spurious duplications, and drops rows due to the lack of matching data or missing values~\cite{joinbugs,joinnull,joinconfusing1}, which are difficult to detect, understand, and ameliorate. We illustrate these with the following example:

\begin{table}[h]
\centering

\begin{minipage}{.13\textwidth}
  \centering
  \begin{small}
    \begin{tabular}{rl}
    \multicolumn{2}{c}{\textbf{\underline{U}}ser} \\
    \toprule
     \textbf{uid} &  \textbf{name} \\
    \midrule
       1 &  Joe \\
       2 &  Mary \\
    \bottomrule
    \end{tabular}
  \end{small}
\end{minipage}%
\begin{minipage}{.15\textwidth}
  \centering
  \begin{small}
    \begin{tabular}{rrr}
    \multicolumn{3}{c}{Purchase \textbf{\underline{H}}istory} \\
    \toprule
     \textbf{uid} &  \textbf{iid} &  \textbf{pid} \\
    \midrule
       1 &    1 &      1 \\
       2 &    1 &      1 \\
       2 &    2 &      \textbf{N} \\
       2 &    4 &      2 \\
    \bottomrule
    \end{tabular}
  \end{small}
\end{minipage}%
\begin{minipage}{.2\textwidth}
  \centering
  \begin{small}
    \begin{tabular}{rrr}
    \multicolumn{3}{c}{\textbf{\underline{I}}tem} \\
    \toprule
     \textbf{iid} &  \textbf{size} &  \textbf{price} \\
    \midrule
       1 &   1 &     20 \\
       2 &   \textbf{N} &     30 \\
       3 &   5 &     35 \\
    \bottomrule
    \end{tabular}
  \end{small}
\end{minipage}%
\vspace{0.2cm}
\begin{minipage}{.15\textwidth}
  \centering
  \begin{small}
    \begin{tabular}{rl}
    \multicolumn{2}{c}{\textbf{\underline{P}}ayment} \\
    \toprule
     \textbf{pid} &  \textbf{name} \\
    \midrule
       1 & Paypal \\
       2 &   Visa \\
    \bottomrule
    \end{tabular}
  \end{small}
\end{minipage}%
\begin{minipage}{.2\textwidth}
  \centering
  \begin{small}
    \begin{tabular}{rrr}
    \multicolumn{3}{c}{\textbf{\underline{A}}d} \\
    \toprule
     \textbf{aid} &  \textbf{source} &  \textbf{cost} \\
    \midrule
        1 &   Google &   500 \\
        2 & Facebook &   600 \\
    \bottomrule
    \end{tabular}
  \end{small}
\end{minipage}%
\begin{minipage}{.12\textwidth}
  \centering
  \begin{small}
    \begin{tabular}{rl}
    \multicolumn{2}{c}{Ad \textbf{\underline{V}}iew} \\
    \toprule
     \textbf{uid} &  \textbf{aid} \\
    \midrule
        1 &    1 \\
       1 &    2 \\
       2 &    1 \\
    \bottomrule
    \end{tabular}
  \end{small}
\end{minipage}%
\vspace{0.2cm}
\caption{Retail Business Database. \textbf{N} is NULL.}
\label{tab:retail}
\vspace{-0.5cm}
\end{table}

\begin{figure}
  \centering
      \includegraphics[width=0.25\textwidth]{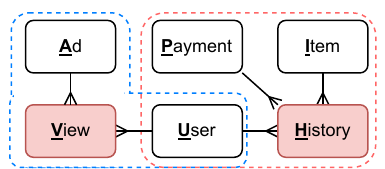}
      \vspace{-0.2cm}
  \caption{
  The join graph for a retail business database. The crow's foot represents the "many" relationships. Within this graph, the fact tables are in \red{red}. The sets of relations to join for denormalization are enclosed within dotted borders.
 }
  \label{fig:retail}
  \vspace{-0.5cm}
\end{figure}

\begin{example}
\label{exp:intro}
The example retail business database \Cref{tab:retail} and join graph \Cref{fig:retail} contain information on the user ad view and purchase history. There are two fact tables: Ad\_view and Purchase.  The standard approach of  denormalization~\cite{denorm,denorm2} can lead to incorrect results.    The denormalized relations are defined as follows:
\begin{lstlisting}
CREATE VIEW Denormalized_Ad_View AS SELECT *
FROM V JOIN U ON V.uid = U.uid 
       JOIN A on V.aid = A.aid;
CREATE VIEW Denormalized_Purchase AS SELECT *
FROM H JOIN U ON H.uid = U.uid 
       JOIN I ON H.iid = I.iid 
       JOIN P ON H.pid = P.pid;
\end{lstlisting}
Let us consider three simple questions and their SQL queries:

\stitle{Q1: What is the total cost of ads from all sources?}
 \begin{lstlisting}
SELECT SUM(A.cost) FROM Denormalized_Ad_View;
\end{lstlisting}
This query above is incorrect because Ad\_view duplicates the cost for each view.  In the table below, Google's cost is double-counted in \red{red}:
\textup{
\vspace*{-.1in}
\begin{table}[h]
\centering
\begin{small}
\begin{tabular}{rrrlrlr}
\textbf{V.uid} &  \textbf{aid} &  \textbf{U.uid} & \textbf{U.name} &  \textbf{A.aid} &   \textbf{source} &  \textbf{cost} \\
   1 &    1 &      1 &  Joe &      1 &   Google &   \red{500} \\
   1 &    2 &      1 &  Joe &      2 & Facebook &   600 \\
   2 &    1 &      2 &  Mary &      1 &   Google &   \red{500} \\
\end{tabular}
\end{small}
\vspace*{-.1in}
\end{table}
}

\noindent We should instead aggregate on only A: \sql{SELECT SUM(A.cost)  FROM A}

\stitle{\bf Q2: What is the total revenue from the purchased items?}
    \begin{lstlisting}
SELECT SUM(I.price) FROM Denormalized_Purchase;
\end{lstlisting}
Unlike ad cost, item price should be duplicated for each purchase in order to compute the {\it total} revenue.  The above query is incorrect because there may be missing payments, and joining on NULL values removes the tuple. One fix is to use outer joins.

\stitle{\bf Q3: What is the total revenue from different ad sources?}
    \begin{lstlisting}
SELECT A.source, SUM(I.price)
FROM V FULL OUTER JOIN U ON V.uid = U.uid 
       FULL OUTER JOIN A on V.aid = A.aid
       FULL OUTER JOIN H ON H.uid = U.uid 
       FULL OUTER JOIN I ON H.iid = I.iid 
       FULL OUTER JOIN P ON H.pid = P.pid
GROUP BY A.source;
\end{lstlisting}
The query above is incorrect because of the one-to-many relationship between a user's revenue (\textbf{Q2}) and their Ad Views; a full outer join would result in duplications and an increase in the total revenue.

One approach widely adopted by the marketing domain~\cite{dbtMarketAtt} is to weigh each ad view based on its ``importance'', and ensure that the sum of weights for each join key value (\sql{uid}) is 1 to counteract join fanouts.  The choice of weights is necessarily based on the analyst's domain knowledge: one analyst may believe that the first ad view is the most important, while another prioritizes the last one~\cite{berman2018beyond}.  The following illustrates weights where all ad views are equally important:

\begin{table}[h]
\vspace*{-.1in}
\centering
\begin{minipage}{.12\textwidth}
  \centering
  \begin{small}
    \begin{tabular}{lr}
    \multicolumn{2}{c}{\textbf{\underline{U}}ser} \\
     \textbf{uid} &  \blue{\textbf{revenue}} \\
    \rowcolor{gray!20}
       1 &  \blue{20} \\
       2 &  \blue{50} \\
    \end{tabular}
  \end{small}
\end{minipage}%
\hspace{0.05cm}
\begin{tikzpicture}[baseline=(current bounding box.center)]
\node[draw, single arrow, minimum height=1.5cm, minimum width=0.2cm, single arrow head extend=0.1cm, fill=white] at (0,0) {};
\node[align=center, above, font=\small] at (0,0.4) {uniformly \\ distributed};
\end{tikzpicture}
\hspace{0.05cm}
\begin{minipage}{.2\textwidth}
  \centering
  \begin{small}
    \begin{tabular}{llrc}
    \multicolumn{4}{c}{Ad \textbf{\underline{V}}iew} \\
     \textbf{uid} &  \textbf{aid} & \blue{\textbf{weight}}  &  \blue{\textbf{revenue}} \\
    \rowcolor{gray!20}
        1 &    1 &   \blue{0.5} & \blue{$20\times0.5$} \\
        \rowcolor{gray!20}
       1 &    2 &   \blue{0.5} & \blue{$20\times0.5$}\\
       2 &    1 &   \blue{1} & \blue{$50\times1$}\\
    \end{tabular}
  \end{small}
\end{minipage}%
\vspace*{-.1in}

\end{table}

For each user, the weights are uniform among ad views. For instance, \sql{uid=1} has weights of 1/2=0.5 as there are 2 views.
Note that the total revenue from User and Ad View are the same (70).
Therefore, analysts can examine how each ad source contributes to revenue based on their assumptions, despite the one-to-many relationship.

Such an idea of "weighing" has been used across multiple domains, as summarized by Kimball and Ross~\cite{kimball2011data}: In order management, freight charges are allocated (or weighed) to a line's products based on their sizes. In financial services, personal incomes are weighed across individual accounts. In accounting, payments are weighed across organizations according to ownership.
\end{example}

From the above example, we see that the correctness of the aggregates  hinges upon the selection of tables to join, deduplication methods, null handling, and weighing designs. Unfortunately, the choices depend on the specific analysis query and the analyst's understanding of the issue at hand, rendering a singular static interpretation unsuitable for all analyses. Nevertheless, the denormalized wide table is still considered the conventional method.

To overcome denormalization's limits and tailor the decisions to specific queries, the industry has developed the notion of a "semantic layer" ~\cite{chatziantoniou2020data,dbtdatamodel,powerbidatamodel,microstrategydatamodel,lookerdatamodel,tableaudatamodel} as a way of decoupling the needs and expertise of data engineers and analysts.
In the offline phase, a data engineer designs a wide table in the form of a {\bf join graph}, along with appropriate {\bf metrics} (i.e., aggregation functions over expressions) pertinent to a given business problem (by discussing the requirements from analysts). In the online exploration phase, analysts use BI tools to specify {\bf exploratory queries} over the metrics with grouping and filtering. Behind the scenes, the semantic layer employs heuristics to determine the tables to join, deduplication methods, and null handling specifically for each query.

\begin{figure}
  \centering
      \includegraphics[width=0.45\textwidth]{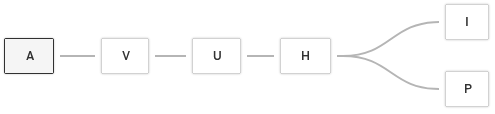}
      \vspace{-0.5cm}
  \caption{Tableau data relationships.}
  \label{fig:tableauJoin}
  \vspace{-0.2cm}
\end{figure}

\begin{figure}
  \centering
  \begin{subfigure}[b]{0.5\textwidth}
    \includegraphics[width=\linewidth]{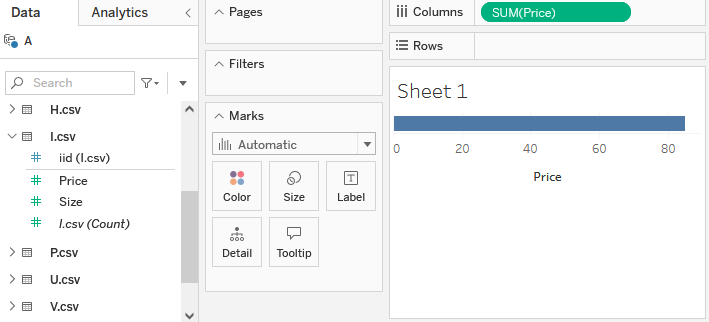}
    \caption{\textbf{Q2}: \sql{SUM(I.prcie)}.}
    \label{fig:tableauQ2}
  \end{subfigure}
  \begin{subfigure}[b]{0.5\textwidth}
    \includegraphics[width=\linewidth]{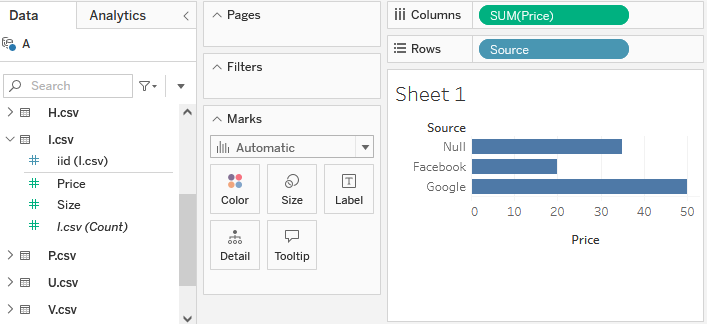}
    \caption{\textbf{Q2}: \sql{SUM(I.prcie) GROUP BY A.source}.}
    \label{fig:tableauQ3}
    \vspace{-0.2cm}
  \end{subfigure}
  \caption{Tableau sheet, where analysts drag the attributes to visualize without worrying about the underlying join relationships. However, Tableau gives wrong results.}
  \label{fig:tableauQ}
  \vspace{-0.5cm}
\end{figure}

For example, Tableau "relationships"~\cite{tableauRelationship} consist of a join graph (\Cref{fig:tableauJoin}), and analysts treat it as a base table and create visualizations by dragging and dropping attributes onto the canvas to explore different metrics, such as \textbf{Q2-3} in \Cref{fig:tableauQ}.  
Unfortunately, to decide the tables to join for each query, Tableau arbitrarily selects those in the minimum sub-tree of the join graph that cover all referenced attributes,
which leads to subtle errors.
For example, \textbf{Q2} only references attribute \sql{I.price}. Tableau, therefore, sums the prices over $I$ and computes the total revenue as \sql{SUM(I.price)=85} (\Cref{fig:tableauQ2}), even though \sql{I.price} should be duplicated for each purchase (by joining with $H$) to compute total revenue correctly.
For \textbf{Q3}, Tableau generates results "Google" $50$ and "FaceBook" $20$, which internally applies specialized deduplication~\cite{tableauRelationship} to handle the join fanout. However, the deduplication mechanism is not surfaced to the analysts, even though it impacts query results and may not align with the analysts' intentions.

Tableau is not alone in making incorrect heuristics. Of the $5$ BI tools and semantic layers we surveyed, 
$2$ of them determine the tables to join in a heuristic manner that's not disclosed to the analyst, which leads to incorrect outcomes for  \textbf{Q2}. Additionally, for \textbf{Q3} with many-to-many relationships, $2$ of these BI tools don't support it, and the remaining $3$ apply arbitrary deduplication rules solely based on heuristics.
They are therefore all susceptible to correctness errors, and hide from the analyst the ability to interpret and control how the final metrics in a query are decided upon.

How do analysts make informed decisions about how to handle duplication problems that arise from joins?   How can analysts best decide to include a table in a query, duplicate or deduplicate tuples, or reweigh duplicates, to correctly compute their desired business metrics? 
This paper formalizes these issues and proposes {\bf ``weighing''} as the core primitive to address them, whose idea is as follows:
Naive query treats every tuple as having an equal "weight" when aggregated. However, with join fanout, the "weights" of join key groups (calculated as the sum of the "weights" of their tuples) can be amplified, which may inadvertently bias the join results.  To overcome this, we  "weigh" tables by assigning equal "weight" to each join group (instead of each tuple) and then distributing the weight among the tuples within the group. Consequently, tuple values are aggregated based on these weights to counteract join fanouts.
Previous BI tools use deduplication, which can be considered as a special "weighing" of $1/0$ without fractional numbers.
We demonstrate how reweighing can resolve fanout-induced errors, and how instances of this formulation have been used in existing problems such as market attribution, unbiased sampling, and ML fairness.

A key challenge that necessitates a human-in-the-loop approach is that the appropriate weighing cannot be determined offline, but is dependent on the specific wide table, query, and even analyst needs.  For instance, the weights for \textbf{Q3} may vary depending on different analysts' belief in the importance of different ad views---Shannon may only care about the ad user viewed last, while Erika  may care about all ad views equally. 
To help users specify the weights and visualize the outcomes, we propose a human-in-the-loop framework. We observe that visualizing each table one-by-one makes it difficult to contextualize the aggregates across joins, and presenting the full join can be overwhelming. Our framework addresses these issues by (1) enabling users to decide the weights iteratively along the join path,  (2) visualizing partial aggregates that summarize the current aggregates for the decided weights (instead of computing the full join), and (3) providing an interface with common weighing options to declaratively specify the weights.

\section{Background}

This section surveys the existing academic literature  for pitfalls that can arise when aggregating over join graphs, 
and identifies the missing gaps and limitations of existing approaches.   
We further survey prominent BI tools for how they address these pitfalls.

\subsection{Pitfalls in Join Aggregation Query}
"Summarizability"~\cite{mazon2009survey,caniupan2012repairing,simon2022controlling} studies the correctness of querying aggregated fine-grained values at a coarser level. Within snowflake-schema multidimensional models, fact values are usually fine-grained, while aggregation queries that join and group dimensional attributes are considered coarse-grained.
To address the fanout issues in \textbf{Q1} and \textbf{Q2}, they require users to specify the "level of detail" of metrics, which determines what should be joined for duplications before the values  are aggregated. They cover other additive issues not directly related to join fanouts. For example, age should not be added, and population can be added over geographical areas but not over time.  These are complementary to this work.

However, "summarizability" is too strict for practical exploratory queries.
For example, when there are missing join keys in dimension tables, "roll up" and "drill down" are considered "incomplete" and therefore not summarizable, which can be addressed by outer join in practice~\cite{tableauRelationship}. Besides, many-to-many joins are considered "non-strict" and also not summarizable, but are important in applications like market attribution~\cite{dbtMarketAtt} and order management~\cite{kimball2011data}. In such cases, weighing can be used (\Cref{sec:solution}) to counteract join fanouts.

The same join-induced issues arise in domains beyond BI. Previous analytics works over joins, such as ML~\cite{schleich2016learning} and sampling~\cite{shanghooshabad2021pgmjoins}, are susceptible to bias~\cite{nargesian2021tailoring} and correctness errors when computed over the materialized join result. Consider the following:

\begin{example}
Consider the database shown in Table \ref{tab:bias}. Suppose analysts aim to train a model to predict user purchases or sample tuples for insights. They want to include features in the join of all three tables $A\Join U \Join H$  for enrichment. Despite having an even gender distribution  in $U$ ($1$ for each), the full join produces  $6\times$ more tuples for female due to fanout. Consequently, this creates a potential imbalance in the ML training and bias for sampled data.
\end{example}
\vspace{-0.3cm}
\begin{table}[h]
\centering
\begin{minipage}{.12\textwidth}
  \centering
  \begin{small}
    \begin{tabular}{rl}
    \multicolumn{2}{c}{Ad \textbf{\underline{V}}iew} \\
      \textbf{aid} & \textbf{uid} \\
        1 &    1 \\
       2 &    2 \\
       2 &    2 \\
    \end{tabular}
  \end{small}
\end{minipage}%
\begin{minipage}{.12\textwidth}
  \centering
  \begin{small}
    \begin{tabular}{rl}
    \multicolumn{2}{c}{\textbf{\underline{U}}ser} \\
     \textbf{uid} &  \textbf{gender} \\
       1  & male \\
       2  & female \\
    \end{tabular}
  \end{small}
\end{minipage}%
\begin{minipage}{.2\textwidth}
  \centering
  \begin{small}
    \begin{tabular}{rr}
    \multicolumn{2}{c}{Purchase \textbf{\underline{H}}istory} \\
     \textbf{uid} &  \textbf{iid}\\
       1 &    1  \\
       2 &    1  \\
       2 &    2  \\
       2 &    4  \\
    \end{tabular}
  \end{small}
\end{minipage}%
\caption{Example databases where gender is evenly distributed in User, but joining it with Purchase results in bias.}
\label{tab:bias}
\vspace{-0.7cm}
\end{table}

A common approach to address data imbalance in the single table regime is via weighing~\cite{ganganwar2012overview}.  We extend weighing to analytics over joins.
Another approach to address join fanout is to pre-aggregate (e.g., averaging)~\cite{frank2007method,guo2008multirelational} to reduce a N-M relationship into a N-1 relationship. However, this introduces errors for long join paths (e.g., the average of averages is not the total average), and so BI tools that apply pre-aggregation only support a single join~\cite{tableauBlend}.

\begin{table}[b]
\centering

\begin{tabular}{llllll}
\textbf{Tool} & \textbf{Method} & \textbf{Source} & \textbf{Q1} & \textbf{Q2} & \textbf{Q3}\\
Tableau & Blend & & $A$  & \red{$I$} & \red{\textbf{NA}}   \\
& Relationship & & $A$ & \red{$I$} & \red{\textbf{WRG}}\\
PowerBI & Relationship & & $A$ & \red{$I$} &  \red{\textbf{ERR}}\\
Looker & Blend & &  $A$ &  \red{$I$} &  \red{\textbf{NA}} \\
&  Model & $V/H$ & \red{$V \leftouterjoin A$} & $H \leftouterjoin I$ & \multirow{2}{*}{\red{\textbf{WRG}}} \\
&  Model & $A/I$ & $A$  & \red{$I$} & \\
Malloy &  Model & $V/H$ &  \red{$V \leftouterjoin A$}  & $H \leftouterjoin I$ & \multirow{2}{*}{\red{\textbf{WRG}}}\\
 & Model & $A/I$ & $A$  & \red{$I$} &  \\
Sigma & Lookup & $V/H$ &  \red{$V \leftouterjoin A$}  & $H \leftouterjoin I$ & \red{\textbf{NA}} \\
&  Lookup & $A/I$ &  $A$   & \red{$I$} & \red{\textbf{NA}} \\
\end{tabular}

\caption{The joins current BI tools use to execute queries.  $\leftouterjoin$ is left outer join. Some methods allow source tables to specify the duplication of measurements.
\red{Red} are wrong results.  \textbf{NA} means the queries can't be specified. \textbf{ERR} means the queries can be specified but errors are thrown. \textbf{WRG} means that the results are based on unintended deduplication mechanisms.}
\label{tab:bitools}
\vspace{-0.5cm}
\end{table}

\subsection{Current Business Intelligence Tools}
\label{sec:currentBI}

Modern BI tools offer different ways to pre-define join graphs and metrics offline, and then allow for online exploratory analysis. For join graphs and metrics, some automatically infer default joins from the database schema, while others offer GUI, SQL, or custom languages for user inputs. 
During online exploration,  analysts query the join graphs as a wide table, 
and the system dynamically determines the appropriate joins to avoid logical pitfalls.    In all cases, they vary in how they handle different fanout conditions (1-1, 1-N, N-M) and NULLs.   We now examine how popular BI tools address the above pitfalls for {\bf Q1-3}.    Our findings are summarized in \Cref{tab:bitools}.

\begin{itemize}[leftmargin=*,itemsep=0pt]
    \item {\bf Blend} is adopted by both Tableau and Looker~\cite{tableauBlend}. 
    The user specifies the join condition between two tables and query, and the BI tool executes it using pre-aggregation before the outer join. Blend is limited in two ways: (1). It only supports queries of join path with a length of 2 (and not general join graphs), therefore doesn't support \textbf{Q3} (with a join path of 5 tables). (2).   
    It doesn't require an explicit definition of the metrics and the decision to join is based on whether the attributes are referenced. For \textbf{Q2}, it avoids the join since the referenced attributes ($I.price$) are not referenced in $H$, which is incorrect because the join is necessary to duplicate prices for the total revenue metric. 

    \item {\bf Relationship} is supported by Tableau and PowerBI, where users specify the join graph (tables and join conditions) and can query over it directly. However, both tools make the same error as Blend for \textbf{Q2} as they don't require explicit metric definitions. For \textbf{Q3}, PowerBI generates errors as it doesn't support many-to-many relationships, while Tableau applies arbitrary deduplication that impacts query results but may deviate from the analyst's intentions without even notifying the analyst.

    \item Looker~\cite{lookerdatamodel} and Malloy~\cite{malloy} use a SQL-based declarative language to {\bf model} a source table to query against, which defines both join graph and metrics. For instance, the following is used for \textbf{Q2}, which defines the join condition with $I$ to construct a join graph, along with the total revenue as metric (\sql{sum(I.price)}).

    \begin{lstlisting}
source: H is table('PurchaseHistory') {
  join_one: I is table('Items') on iid = I.iid
  measure: total_revenue is sum(I.price)
}
    \end{lstlisting}
    Note that there are different options for selecting the source table. In the example, $H$ is the source, but $I$ could also be the source. The choice of the source table determines the  metric duplication level: In this example, because the total revenue metric is sourced from $H$, it is aggregated after $H \leftouterjoin I$.
    For \textbf{Q1}  and \textbf{Q2}, the correctness depends on if the correct source table  is used for duplication ($A$ for \textbf{Q1} and $H$ for \textbf{Q2} in the \textbf{Source} column). For \textbf{Q3}, choosing different source tables yields different results, and both BI tools apply arbitrary deduplication without user interactions. 
    
    \item Finally, Sigma Computing supports queries over attributes across two tables through {\bf lookup} join~\cite{sigmaLookup}, which employs preaggregations and is not applicable to \textbf{Q3} similar to {\bf blend}. Unlike {\bf blend} but similar to {\bf model}, users can specify explicitly whether the aggregation is being performed before or after join for \textbf{Q1-2}.

\end{itemize}

In summary, current BI tools frequently depend on implicit assumptions for deduplication, which can be dangerous if they are incorrect and challenging to identify the errors.

\section{weighing to the Rescue}
In this work, we argue that users should be able to easily specify reweighing policies in order to avoid correctness errors when performing analytics over joins.
This section first defines the consistency errors in the context of semantic layers that store pre-defined join graphs and associated metrics.   Our solution then builds on semi-ring aggregation.   Although semi-ring aggregation is traditionally used to accelerate join-aggregation queries by pushing the aggregations through joins, this push-down computes partial aggregates that are also useful for reasons about the appropriate reweighing decisions.

\begin{table*}[h]
\centering
\begin{minipage}{.2\textwidth}
  \centering
  \begin{small}
    \begin{tabular}{rrr>{\columncolor[gray]{0.8}}c}
    \multicolumn{4}{c}{$H$} \\
    \toprule
     \textbf{uid} &  \textbf{iid} &  \textbf{pid} & \textbf{Ann}\\
    \midrule
       1 &    1 &      1 &      1\\
       2 &    1 &      1 &      1\\
       2 &    2 &      \textbf{N} &      1\\
       2 &    4 &      2 &      1\\
    \bottomrule
    \end{tabular}
  \end{small}
\end{minipage}%
\hspace{0.05cm}
\begin{tikzpicture}[baseline=(current bounding box.center)]
\node[draw, single arrow, minimum height=1.5cm, minimum width=0.2cm, single arrow head extend=0.1cm, fill=white] at (0,0) {};
\node[align=center, above, font=\small] at (0,0.4) {partial  \\ aggregate};
\end{tikzpicture}
\hspace{0.05cm}
\begin{minipage}{.12\textwidth}
  \centering
  \begin{small}
    \begin{tabular}{r>{\columncolor[gray]{0.8}}c}
    \multicolumn{2}{c}{$\gamma_{uid}H$} \\
    \toprule
     \textbf{uid} &  \textbf{Ann}\\
    \midrule
       1 &     1\\
       2 &     3\\
    \bottomrule
    \end{tabular}
  \end{small}
\end{minipage}%
\begin{tikzpicture}[baseline=(current bounding box.center)]
\draw[->,thick] (0,0) -- (0.5,0);
\end{tikzpicture}
\hspace{0.05cm}
\begin{minipage}{.12\textwidth}
  \centering
  \begin{small}
    \begin{tabular}{r>{\columncolor[gray]{0.8}}c}
    \multicolumn{2}{c}{$\gamma_{uid}H \leftouterjoin\gamma_{uid}V$} \\
    \toprule
     \textbf{uid} &  \textbf{Ann}\\
    \midrule
       1 &     $1\times2$\\
       2 &     $3\times1$\\
    \bottomrule
    \end{tabular}
  \end{small}
\end{minipage}%
\hspace{0.2cm}
\begin{tikzpicture}[baseline=(current bounding box.center)]
\draw[->,thick] (0.5,0) -- (0,0);
\end{tikzpicture}
\begin{minipage}{.12\textwidth}
  \centering
  \begin{small}
    \begin{tabular}{r>{\columncolor[gray]{0.8}}c}
    \multicolumn{2}{c}{$\gamma_{uid}V$} \\
    \toprule
     \textbf{uid} &  \textbf{Ann}\\
    \midrule
       1 &     2\\
       2 &     1\\
    \bottomrule
    \end{tabular}
  \end{small}
\end{minipage}%
\hspace{0.05cm}
\begin{tikzpicture}[baseline=(current bounding box.center)]
\node[draw, single arrow, minimum height=1.5cm, minimum width=0.2cm, single arrow head extend=0.1cm, fill=white, rotate=180] at (0,0) {};
\node[align=center, above, font=\small] at (0,0.4) {partial  \\ aggregate};
\end{tikzpicture}
\hspace{0.05cm}
\begin{minipage}{.12\textwidth}
  \centering
  \begin{small}
    \begin{tabular}{rl>{\columncolor[gray]{0.8}}c}
    \multicolumn{3}{c}{$V$} \\
    \toprule
     \textbf{uid} &  \textbf{aid} &  \textbf{ann}\\
    \midrule
        1 &    1 &    1 \\
       1 &    2 &    1 \\
       2 &    1 &    1 \\
    \bottomrule
    \end{tabular}
  \end{small}
\end{minipage}%
\vspace{0.2cm}
\caption{Partial aggregations $\gamma(H \leftouterjoin V) = \gamma(\gamma_{uid}H \leftouterjoin \gamma_{uid}V)$ reduce many-to-many joins to one-to-many joins.}
\label{tab:partialagg}
\vspace{-0.7cm}
\end{table*}

\subsection{Problem}
\label{sec:prob}

We consider the setting where a data engineer has pre-defined (1) the metric as an aggregation function and (2) the duplication as an acyclic join; the aggregation and join together constitute a {\it base query} $Q_{base}=\gamma(R_1\leftouterjoin\cdots\leftouterjoin R_n)$ (without group-by and selection). 
For example, the base query for \textbf{Q2} can be represented as $Q_{base}=\gamma_{SUM(I.price)}(H \leftouterjoin I)$.
The analyst then composes an SPJA query $Q$ that uses the same metric but may also include selection and group-by expressions that reference attributes in tables that require left outer joining with additional tables.   
For instance, to answer \textbf{Q3}, they may issue $Q=\gamma_{\red{A.source},SUM(I.price)}(H \leftouterjoin I \red{\leftouterjoin U \leftouterjoin V \leftouterjoin A})$.

We next formalize the "consistency" error.
To facilitate understanding, we consider "sum" as the aggregation in this section, and we will extend it to arbitrary aggregation when introducing the solution framework.
The {\it "consistency"  error} refers to 
the inequality  between the sum of the groupby \footnote{Selection removes data and the inequality is expected. For the problem definition purposes, we regard "selection" as a groupby based on whether the selection predicate is satisfied and post-process it to obtain the final selected value.} results for $Q$ and $Q_{base}$:
 $Q$ includes additional joins for enriched group-bys and selections, and the total measurement (e.g., revenue or expense) can be amplified compared to $Q_{base}$. Such inconsistently larger results have been complained about  across different BI tools~\cite{joinconfusing1,joinconfusing2,joinconfusing3,joinconfusing4,joinconfusing5,joinconfusingtweet3,joinconfusingtweet2}, and we want to help analysts understand and avoid them.

Given the base query $Q_{base}=\gamma_{SUM}(R_1\leftouterjoin...R_k)$, and the exploration query $Q=\gamma_{\red{gb},SUM}(\red{\sigma}(R_1\leftouterjoin...R_k\red{\leftouterjoin...R_j}))$, that includes additional joins, selections, and group-bys (in red), the objective is to find the re-weighing function $W$ that weighs (multiplies) the sum when joined (e.g., the revenue is weighed for \textbf{Q3} in \Cref{exp:intro}), such that $Q^*=\gamma_{gb}(\sigma(R_1\leftouterjoin...R_k\leftouterjoin\blue{W}(R_{k+1})...\blue{W}(R_j)))$ is:

\begin{itemize}[leftmargin=*,itemsep=0pt]
    \item {\bf Consistent}: 
    If $Q$ doesn't have a selection $\sigma$, then $\gamma(Q^*)= Q_{base}$.
    If $Q$ has a selection $\sigma$, let $\neg \sigma$ be its negation and $\gamma(Q^*_\neg)=\gamma_{gb}(\red{\neg}\sigma(W(R_1)\leftouterjoin...W(R_j)))$.
    Then $\gamma(Q^*) + \gamma(Q^*_\neg) = Q_{base}$.

    \item {\bf User-Directed}: There could be various valid weighing strategies. For example, in \textbf{Q3}, different analysts may assign different weights based on their opinions about the importance of the ad views. The weighing should be transparent and understandable to analysts, who should be able to guide based on their interpretation of the data and domain knowledge.
\end{itemize}   

We observe that the re-weighing function \blue{$W$} is applied to relations only in the exploratory query ($R_{j+1}\leftouterjoin...R_k$) to mitigate fanout effects, but not to those in the base query. The duplication in the base query is assumed intentional to compute the total metric correctly. Although there might be other reasons to weigh the base query relations, such as addressing an imbalanced distribution~\cite{nargesian2021tailoring}, we consider these as a preprocessing step, separate from \blue{$W$}.

\stitle{Scope.} Many previous works~\cite{kearns1997teaching,grust2011true,presler2021sqlrepair,ahadi2016students} studied syntax and semantic issues related to SQL queries for correctness, which remains challenging. We specifically focus on the amplified or reduced aggregate result caused by the join fanout and ensure "consistency" as formally defined above.  We hope that, ensuring consistency helps users identify semantic errors and achieve final query correctness.

\subsection{Semi-ring Aggregation Preliminary}

Semi-ring aggregation breaks down the aggregation into two fundamental operations: addition (e.g., to aggregate values) and multiplication (e.g., to weigh values).  It is highly expressive and can express nearly all common aggregations with the benefit of efficient partial aggregation, such as sum, count, average, and max. Other aggregations, such as median, can also be expressed using semi-ring aggregation (by exhaustively tracking all values in a long list), but don't derive as much benefit from partial aggregation for both performance and interpretability perspectives.

\stitle{Data Model.} 
We use the traditional relational data model: Given relation $R$, let $A$ be an attribute, $dom(A)$ be its domain, $S_R=[A_1,\cdots,A_n]$ be its schema, $t\in R$ be a tuple of $R$, and $t[A]$ be the value of attribute $A$ in tuple t. The domain of $R$ is then the Cartesian product of attribute domains, i.e., $dom(R) = dom(A_1)\times\cdots\times dom(A_n)$.

\stitle{Semi-ring Aggregation Query.} We begin by extending the relation table with annotations,\cite{green2007provenance,joglekar2015aggregations,nikolic2018incremental}, which maps $t\in R$ to a commutative semi-ring $(D, +, \times, 0, 1)$, where $D$ is a set, $+$ and $\times$ are commutative binary operators closed over $D$, and $0/1$ are the zero/unit elements. Annotations are useful for query optimizations based on algebraic manipulation. Different semi-ring definitions support various aggregation functions, from standard statistical functions to machine learning models. For example, the natural numbers semi-ring $(\mathbb{N},+,\times,0,1)$ allows for integer addition and multiplication, and supports the \cnt aggregate. For an annotated relation $R$, let $R(t)$ represent the annotation of tuple $t$. 
Tuples in the domain but not in the table are assumed to have 0 as annotations.

Aggregation queries can now be redefined over annotated relations by translating group-by and join operations into $+$ and $\times$ operations over the semi-ring annotations, respectively:

\begin{align}
  (\gamma_\mathbf{A} R)(t) = & \sum \{R(t_1) | \HS t_ 1 \in R , t = \pi_{\mathbf{A}} (t_1 )\} \\
(R\Join T)(t) =& \HS R(\pi_{S_R} (t)) \times T(\pi_{S_T} (t)) 
\end{align}
\noindent (1) The annotation for each group-by result in $\gamma_\mathbf{A} R$  is the sum of the annotations of all tuples in its input group. (2) The annotation for each join result in $R \Join T$ is the product of semi-ring annotations from its contributing tuples in $R$ and $T$.  The $\Join$ can be extended for outer join, where the non-matching tuple retains its original annotation and has the rest of the attributes as NULL.

To translate an aggregation function like SUM into semi-ring aggregation, we need to determine (1) the semi-ring and (2) the annotation for initial tables. For example, in the case of \sql{SUM(I.price)}, the semi-ring is the real number following standard mathematics. For the annotation, only $I$ has each tuple $t$ annotated with \sql{t[price]}; all other tables have all annotations of 1.

\stitle{Partial Aggregation.}
The key optimization in factorized query execution~\cite{abo2016faq,schleich2016learning} is to distribute aggregations (additions) through joins (multiplications). 
Consider the $\gamma(H \leftouterjoin V)$ for count semi-ring, which is a many-to-many join. We can push down part of aggregations before join for one-to-many as illustrated in \Cref{tab:partialagg}.

\subsection{Consistency through Weighing}
\label{sec:solution}
There are many aggregations like min/max~\cite{mazon2009survey} and other specialized probabilistic databases~\cite{koller2009probabilistic,abo2016faq} that provides consistent results, even with many-to-many joins. We highlight the key property they have that ensures consistency, and generalize it to other aggregations through reweighing.

Given $Q_{base}=\gamma(R_1\leftouterjoin...R_k)$, consider the query with an additional join $\gamma(Q)=\gamma(R_1\leftouterjoin...R_k  \leftouterjoin \red{R_{k+1}}) = \gamma(\gamma_{J}(R_1\leftouterjoin...R_k) \leftouterjoin \red{\gamma_{J}(R_{k+1})})$, where $J$ is the join key between $R_{k+1}$ and the rest of the relations.
Then, the key sufficient condition for consistency is that: 
$$\forall j\in Dom(J), \gamma_{J}(R_{k+1})(j) = 1$$ 
The consistency $\gamma(Q) = Q_{base}$ is ensured because of the multiplicative identity property of $1$ element in semi-ring, and can be recursively applied to an arbitrary number of joins. This is also satisfied with additional group-bys (because the groupings are summed) and selections (as special group-bys of whether selected or not).

Previous applications that support many-to-many joins satisfy this property. For probabilistic tables, it is ensured that each conditional probability table has the sum of the probabilities conditioned (selected) on the join key as $1$. In the case of min/max aggregation, the relation without the attribute to maximize have all tuples annotated with $1$, and the min/max semi-ring has its addition  operator $\bigoplus$ as $max$ that ensures $1 \bigoplus ... \bigoplus 1 = max(1,..., 1) =1$.

For other aggregation queries such as sum and count, such a property is not natively satisfied and it's necessary to weigh relations. For instance, in the context of market attribution of \textbf{Q3}, we can assign equal weights to all tuples in $V$ with the same \sql{uid}. However, assigning weights is not one-time, as different analysts may have different opinions on how to weigh the ad views. Therefore, we next introduce a framework to assist users in assigning weights.

\begin{figure}
  \centering
      \includegraphics[width=0.42\textwidth]{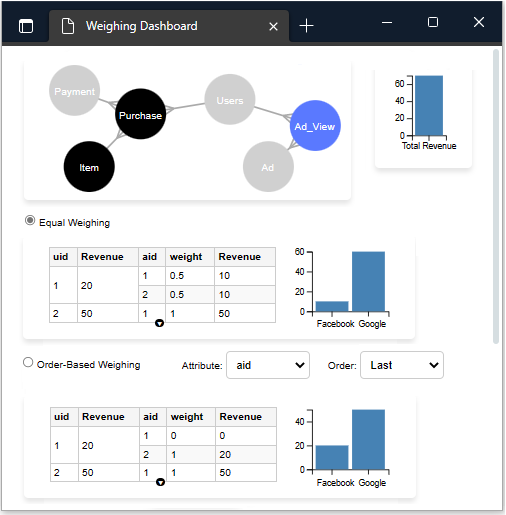}
      \vspace{-0.2cm}
  \caption{The weighing interface screenshot displays the join graph and base query result visualizations at the top. The tabular view of the weighed relations and visualizations are displayed for different  weighing options.
 }
  \label{fig:interface}
  \vspace{-0.6cm}
\end{figure}

\subsection{Human-in-the-loop Weighing}

To make informed decisions, users need to contextualize aggregation within the larger join graph and be capable of specifying weights for the joined table. However, visualizing each table individually makes it challenging to contextualize aggregates across joins, while presenting the full join can be overwhelming.
A key design decision for understanding the aggregation relationship is utilizing partial aggregates. These partial aggregates progressively join with new tables, compress the results by aggregating out attributes, not in use and retaining only the necessary attributes for users to comprehend and design weights.
The challenge in specifying weights lies in the worst-case scenario, where users need to assign a weight for each tuple, which is impractical. Instead, we identify common and efficient special cases that simplify the process for the user. We aim to determine the minimum information required, given reasonable assumptions for specific settings that are clearly stated and easier for the analyst to specify. This section delves into these details and introduces an interface.

\stitle{Interface.} Our interface comprises two panels illustrated in \Cref{fig:interface} to solicit weighings from users, and visualize outcomes.

\subsubsection{Weighing.} 
For exploratory query $Q$, we  request weighing only the necessary relations in a depth-first fashion, where the relations in $Q_{base}$ are the roots.
Some relations, such as those in the $Q_{base}$ don't require weighing (\Cref{sec:prob}).  
Other relations, with one-to-many or one-to-one relationships (to the parent relation in the depth-first search), have one tuple per join key and are directly annotated with a weight of 1 without user request. We only request weights to relations that (1) have  many-to-one or many-to-many relationships and (2) are part of any paths from the $Q_{base}$ relations to the $Q$ relations, excluding the $Q_{base}$ relations themselves.

Based on the sophistication of users, we offer parameterized options for common cases to streamline the process such as:

\begin{itemize}[leftmargin=*,itemsep=0pt]
    \item {\it Equal Weighing}: Tuples within the group have the same weight.
    \item {\it Order-Based Weighing}: The first/last tuple ordered by some user-specified attribute has weight 1, while the rest have 0.

    \item {\it Position-Based Weighing}: Similar to order-based one, but users can additionally determine the allocation percentages for the first/last tuple and the rest. E.g., the first and last have weights of 0.4, while the remaining 0.2 is distributed evenly for the rest.

    \item {\it Proportional Weighing}: The weights are distributed proportionally to some user-specified attribute (e.g., distributing freight charges based on the item sizes~\cite{kimball2011data}).
    
\end{itemize}

For advanced users, we provide a SQL-based interface that allows for customized weighing specifications.
The SQL query is intended to create a weight column. Since relations are unordered, we ask users to create a weight table $W[rowid, weight]$ and then join it with the relation to append the weight column. 
For instance, the uniform weighing (linear attribution) in \Cref{exp:intro} \textbf{Q3} can be specified using the following SQL queries:

\begin{lstlisting}
SELECT rowid, 1/COUNT(*) 
       OVER (PARTITION BY uid) AS weight
FROM V;
\end{lstlisting}
Users can then modify SQL queries to customize the weights as per their requirements. The weighing can also be based on multiple attributes and even other tables in SQL. Once the weight column is specified, we perform sanity checks: For exploratory queries, we verify whether the weights grouped by the join key are all $1$.

\subsubsection{Visualization.} We have developed two types of visualizations for the weighing interface: (1) Tabular views of weights for detailed weights, (2) Visualizations of query results (e.g., $Q_{base}, Q$) over the whole join along with the join graph for an overview view.
During the weighing process, we provide users with a tabular view of the relation with weights for  detailed inspection. 
It is necessary to visualize both the relation being weighed and the relation it joins, to understand the distribution of the aggregation. But these two relations have potentially many-to-many relationships, which is hard to visualize.
To address this, we implement partial aggregation to summarize aggregates of the parent table from the depth-first search grouped by its join key  and present it as a one-to-many join using nested table layout~\cite{bakke2016expressive}. For instance, in \Cref{fig:interface}, we aggregate the revenue of each user by the join key "uid". The relation could be large, and we only sample $n$ ($=100$ by default) join key groups to display, but the table can be expanded to display more rows (by clicking on the button at the bottom of the table).

For visualizations, we display  $Q_{base}$ and $Q$ by default but offer a library for users to build extra visualizations. 
Before weighing, we use the "equal weighing" by default to compute the query results for visualizations, which are progressively updated as users specify weighing. 
We place $Q_{base}$ at the top since it remains unaffected by weighing and show other possible visualizations for various weighing options to assist users in understanding potential outcomes. Additionally, we present the join graph, which depicts the relationships (e.g., many-to-many, one-to-one), highlighting the $Q_{base}$ relations (in black), the relation being weighed (in blue).

\subsection{Limitations and Future works}
Our existing interface evaluates each metric independently, necessitating users to examine and navigate various metrics individually. However, users might prefer to compare and visualize multiple metrics collectively and further generate metrics based on the existing ones. In our future work, we plan to create a unified interface that not only displays multiple metrics but also enables users to seamlessly comprehend weighed outcomes across diverse metrics and weighing options, in order to fully actualize the semantic layer.

\pagebreak

\bibliographystyle{ACM-Reference-Format}
\bibliography{main}

\clearpage

\end{document}